\documentclass{aa}
\usepackage[switch]{lineno}
\usepackage{natbib,twoopt}
\bibpunct{(}{)}{;}{a}{}{,} 
\usepackage[breaklinks=true]{hyperref}
\usepackage{graphicx}
\usepackage{pifont}
\usepackage{txfonts}
\usepackage{xcolor}

\begin{document}
\begin{linenumbers}
   \title{Can diffuse X-rays be important in driving photoionisation in molecular clouds?}

   \authorrunning{E. I. Makarenko et al.}
   \author{E. I. Makarenko\inst{1}\thanks{emak@mpe.mpg.de},
           A. V. Ivlev\inst{1},
           S. Bialy\inst{2},
           X. Zheng\inst{1},
           B. A. L. Gaches\inst{3}, 
           P. Caselli\inst{1},
           M. Padovani\inst{4},
           M. C. H. Yeung\inst{1}
          }

   \institute{Max Planck Institute for Extraterrestrial Physics, Giessenbachstrasse 1, 85748 Garching, Germany\and
   Physics Department, Technion - Israel Institute of Technology, Haifa 3200003, Israel\and
    Faculty of Physics, University of Duisburg-Essen, Lotharstr. 1, Duisburg 47057, Germany \and
    INAF–Osservatorio Astrofisico di Arcetri, Largo E. Fermi 5, 50125 Firenze, Italy
              }

   \date{Received 5 August 2026 / Accepted 1 September 2026 }
 
  \abstract
   {The ionisation balance in molecular clouds is regulated by several ionising sources, including cosmic rays, X-rays, and ultraviolet radiation. Their relative importance depends on the local physical conditions and on the shielding column density.}
   {We compute the contribution of the large-scale diffuse X-ray radiation field to the ionisation in molecular clouds in the absence of strong local X-ray sources, such as young stars. Our goal is to quantify its significance relative to the Galactic cosmic rays.}
   {Using measurements of diffuse X-ray emission from the first \textit{eROSITA} all-sky survey, we estimate the depth-dependent X-ray ionisation rate in molecular clouds across different Galactic environments. We consider both the observed diffuse field and a deabsorbed field obtained by correcting for foreground Galactic attenuation using HI4PI line-of-sight column densities in a pixel-by-pixel framework.}
  {Diffuse Galactic X-rays are unlikely to dominate the ionisation balance in well-shielded molecular gas ($N_{\mathrm{H}} > 5 \times 10^{21} \mathrm{cm^{-2}}$), where cosmic rays remain the primary ionising agent.
  At low column densities, however, the deabsorbed diffuse X-ray field produces ionisation rate of $\zeta_X \sim 10^{-18}\mathrm{s^{-1}}$ at $N_{\rm H} \sim 10^{19}\mathrm{cm^{-2}}$, remaining below the lower end of Galactic cosmic ray ionisation rate ($\sim 10^{-17} \mathrm{s^{-1}}$). The X-ray contribution decreases rapidly with increasing shielding, falling to $\zeta_{X}\sim10^{-19}\mathrm{s^{-1}}$ by $N_{\rm H}\sim10^{21}\mathrm{cm^{-2}}$ and to   $\zeta_{X}\sim5\times10^{-21}\mathrm{s^{-1}}$ by $N_{\rm H}\sim10^{22}\mathrm{cm^{-2}}$. Diffuse X-ray may therefore contribute to the thermal and chemical structure of low-extinction material and should be considered when interpreting ionisation tracers in diffuse or cloud-envelope environments. 
  }
   {}


   \maketitle
\nolinenumbers

\section{Introduction}

The thermal and chemical state of molecular clouds (MCs) is regulated by external radiation fields. 
In the outer layers of MCs, the dominant driver is typically far-ultraviolet (FUV) radiation, which produces the classical photon-dominated region (PDR) structure \citep{Hollenbach1999, Tielens2005}. 
However, FUV photons are quickly absorbed by dust and molecular self-shielding, and their impact rapidly declines beyond column densities of  $N{_\mathrm{H}}\sim10^{19}$--$10^{21} \ \mathrm{cm^{-2}}$, depending on the FUV intensity, gas density, and dust properties \citep{Tielens1985, Sternberg1995, Hollenbach1999, Bialy2016}. 
Beyond this, the chemistry and thermal balance of the gas are generally regulated by ionising particles that can penetrate deeper in the gas, mainly low-energy cosmic rays (CRs; E $<1$~GeV) \citep[e.g.][]{Dalgano2006, Padovani2020, Gaches2026}.
Therefore, numerous observational studies have used molecular ions such as H$_3^+$, OH$^+$, and H$_2$O$^+$, as well as H$_2$ rovibrational emission, to constrain CR ionisation rates (CRIR) across environments with column densities from around~$10^{19}$ to $10^{23}\,\mathrm{cm^{-2}}$ \citep[e.g.][]{Neufeld2017, Bialy2022, Bialy2026, Indriolo2026}. 
These studies typically derive CRIRs in the range $\zeta\sim10^{-17}$--$10^{-16},\mathrm{s^{-1}}$, with substantial environmental variation.

On the other hand, X-rays can also penetrate gas shielded from UV radiation and can produce chemical and excitation signatures commonly associated with CR ionisation, including molecular ions and H$_2$ excitation \citep{Maloney1996, Jenkins2013, Wolfire2022}.
If diffuse X-ray fluxes are sufficiently high, part of the ionisation in observations could originate from X-rays rather than CRs, potentially affecting the interpretation of CRIRs in the interstellar medium (ISM).
Diffuse X-ray photons are, however, expected to attenuate more strongly with column density than CRs. 
Soft X-rays (E $\approx 0.2$--$1$ keV) are absorbed at columns of $N_{\mathrm{H}}\sim10^{20}$--$10^{21}\ \mathrm{cm^{-2}}$, while harder photons can penetrate to a few $\times10^{22}\ \mathrm{cm^{-2}}$ \citep{Cruddace1974, Maloney1996, Wilms2000, Meijerink2005, Langer2015}.
X-ray shadow studies toward giant MCs and the Eos cloud further indicate that diffuse X-ray backgrounds are largely excluded from MC interiors \citep{Snowden1990, Snowden2000,Yeung2023, Burkhart2025}.
Therefore, the main open question is whether the diffuse interstellar X-ray radiation field can contribute significantly to ionisation, for example, in the outer layers of MCs.

In addition to diffuse backgrounds, X-ray emission from discrete sources (in particular young stellar objects) has been extensively studied in nearby star-forming regions such as Taurus.
Observations from the XMM-Newton Extended Survey of Taurus (XEST) and complementary multi-wavelength programmes show that pre-main-sequence stars produce frequent and luminous X-ray emission capable of affecting the ionisation and chemistry of surrounding gas \citep[e.g.][]{Guedel2007}.
Theoretical models of X-ray dominated regions (XDRs) further demonstrate how keV photons from compact sources heat and ionise dense gas, complementing classical PDRs \citep{Meijerink2005}. 
In protoplanetary discs, stellar X-rays can regulate the ionisation fraction in surface layers and winds, while midplanes remain largely shielded \citep{Glassgold2012}.
The concept of “Röntgen spheres” has also been introduced to describe the localised regions of enhanced ionisation around active stars \citep{Locci2018}. 
These works show that stellar X-rays can have a significant impact, but mainly on sub-parsec to parsec scales in the immediate vicinity of sources. 
However, the present work addresses the large-scale diffuse X-ray field that is present everywhere and its cumulative effect on MCs. 
Such radiation is more spatially extended but generally weaker and more strongly attenuated.
For example, X-ray surveys of massive star-forming regions in the Milky Way and nearby galaxies have shown that MCs are bathed in X-ray emission, 
especially in extreme environments such as 30 Doradus \citep{Feigelson2007, Townsley2011, Townsley2014}.

In this work, we use diffuse \textit{eROSITA} all-sky survey measurements to calculate the X-ray ionisation rate across representative Galactic environments and determine the minimum diffuse X-ray component required to dominate over the Galactic CRIR range.
We find that the X-ray ionisation rate decreases with column density much more steeply than the CRIR.
As a result, diffuse X-rays become orders of magnitude weaker than CR ionisation at large column densities, and even at low columns, X-ray ionisation remains generally below canonical CR ionisation rates.
The paper is structured as follows: Sect.~\ref{sec:method} describes the data and methodology, Sect.~\ref{sec:results} presents the results, Sect.~\ref{sec:discussion} discusses the associated uncertainties, and Sect.~\ref{sec:conclusions} summarises our conclusions.

\section{Method}\label{sec:method}
\subsection{Deriving the X-ray intensity from eROSITA}

\begin{table}
\caption{Galactic regions in eRASS1.}
\centering
\begin{tabular}{lcc}
\hline\hline
Region & $l$ (deg) & $b$ (deg) \\
\hline
I (Disc) & [180, 360] & [-15, +15] \\ 
II (outflow N) & [270, 360] & [+15, +90] \\ 
III (outflow S) & [270, 360] & [-90, -15] \\ 
IV (anti-centre N) & [180, 270] & [+15, +90] \\ 
V (anti-centre S) & [180, 270] & [-90, -15] \\ 
\hline
\label{tab:gal_regions}
\end{tabular}
\end{table}

We use diffuse X-ray maps from the first \textit{eROSITA} all-sky survey (eRASS1), with resolved point sources removed down to a conservative flux threshold (e.g. $10^{-13}$\,erg\,s$^{-1}$\,cm$^{-2}$). This ensures that the remaining emission is dominated by large-scale diffuse components, although residual contamination from unresolved sources cannot be fully excluded.
Following \citet{Zheng2024}, we define five representative Galactic regions (disc, northern and southern outflow, and northern and southern anti-centre; see Table~\ref{tab:gal_regions} for details) and extract the mean diffuse surface brightness in each energy band from $0.2$ to $5.0$~keV.
These regions are chosen to sample different Galactic environments with varying column densities and levels of diffuse emission.
Photons below 0.2~keV may provide an additional ionisation in low-column-density gas.
However, this represents an extreme-ultraviolet contribution outside the X-ray energy range considered in this work and is expected to be quickly attenuated in neutral gas.

The input maps provide count rates in units of [counts\,s$^{-1}$\,pixel$^{-1}$], with a pixel size of $3'\times3'$ (corresponding to an area of $9$\,arcmin$^2$). These are first converted to count-rate surface brightness in units of [counts\,s$^{-1}$\,deg$^{-2}$] by dividing by the pixel area. 
A constant particle background (filter-wheel-closed data or FWC, expressed in the same units) is then subtracted.
The resulting count-rate surface brightness is converted to energy surface brightness using published energy conversion factors (see Table~2 of \cite{Zheng2024}).

For each Galactic region and energy band, we compute the pixel-by-pixel surface brightness values:
\begin{equation}
    I_{\rm band} = \frac{1}{N_{\rm pix}}\sum_i I_{{\rm band},i},
\end{equation}
where $I_{{\rm band},i}$ is the surface brightness of pixel $i$ in units of $[\mathrm{erg \ s^{-1} \ cm^{-2} \ deg^{-2}}]$, and  $N_{\rm pix}$ is the number of spatial pixels within the selected region.

To estimate the radiation incident on MC, each pixel surface brightness is converted from units of [erg s$^{-1}$ cm$^{-2}$ deg$^{-2}$] to specific intensity per steradian [erg s$^{-1}$ cm$^{-2}$ sr$^{-1}$]. 

We describe the incident radiation field in terms of the mean intensity $J$, instead of a radiation flux. 
For an isotropic radiation field, $J=I$ and the angle integrated intensity is $4\pi J = 4\pi I$.
At the surface of an optically thick MC, however, only the incident hemisphere contributes to the radiation field within the cloud.
The mean intensity at the cloud surface is therefore $J = I/2$, such that for pixel $i$:
\begin{equation}
4\pi J_{{\rm band},i}=2\pi I_{{\rm band},i}.
\end{equation}
The corresponding regional mean angle-integrated intensity is:
\begin{equation}
4\pi J_{\rm band} =
\frac{1}{N_{\rm pix}}\sum_i 4\pi J_{{\rm band},i}.
\end{equation}

\subsection{Calculating the X-ray ionisation rate}
To reconstruct the differential angle-integrated intensity within each band, we assume a power-law spectrum:
\begin{equation}
4\pi J_E(E) = KE^{1-\Gamma},
\end{equation}

with photon index $\Gamma = 2$ (for the  diffuse soft X-ray background using eROSITA $\Gamma$ value see \cite{Yeung2024}; and see Sect.~\ref{sec:discussion_index} for a discussion on the impact of $\Gamma$ for our calculations).

The normalisation $K$ is set by requiring that the band-integrated angle-integrated intensity reproduces the deabsorbed band radiation field associated with a given line-of-sight (LOS) column (or, equivalently, an $N_{\rm H,LOS}$ bin in the pixel-wise deabsorption procedure), i.e.
\begin{equation}
K = \frac{4\pi J_{\rm band}} {\int_{E_{\min}}^{E_{\max}}E^{1-\Gamma}\mathrm{d}E},
\end{equation}

where $E_{\min}$ and $E_{\max}$ define the limits of the selected energy band.

We define $\zeta_{\rm X}(N_{\rm H})$ as the effective X-ray ionisation rate per H nucleus at column density $N_{\rm H}$ and compute it as:
\begin{equation}
\zeta_{X}(N_{\rm H}) = \int_{E_{\min}}^{E_{\max}} \frac{4\pi J_E(E)}{E} \sigma_{\rm abs}(E) \exp[-\sigma_{\rm abs}(E)N_{\rm H}] Y(E)\mathrm{d}E,
\end{equation}
where $J_E(E)$ is the differential mean intensity of the radiation field at the cloud surface, $\sigma_{\rm abs}(E)$ is the effective photoabsorption cross-section per H nucleus \citep{Verner1995, Longair2011}, and $Y(E)$ is the total number of ionisations produced per absorbed photon.
The calculated $\sigma_{\rm abs}(E)$ is taken from \cite{Gaches2023}\footnote{https://github.com/AstroBrandt/XRayCrossSections}, assuming Solar abundances and a predominantly neutral medium.
In the 0.2--5~keV range, the $\sigma_{\rm abs}(E)$ decreases approximately as $\sigma_{\rm abs}(E)\propto E^{-3}$ between absorption edges.

Each primary photoelectron produces secondary ionisations as it slows down in the gas.
At X-ray energies, this process is highly efficient because the large excess energy of the photoelectrons allows them to undergo multiple inelastic collisions, each of which is capable of producing additional ionisation.
We include this contribution using:
\begin{equation}
Y_{\rm sec}(E) = f_{\rm ion}(x_e)\, \frac{E - 13.6\,{\rm eV}}{W},
\end{equation}
where $W \simeq 36$~eV is the mean energy required to produce one ion pair in predominantly neutral gas, and $f_{\rm ion}(x_e)$ is the fraction of the electron energy deposited into ionisation, which decreases with increasing free-electron fraction $x_e$ \citep{Shull1985,Dalgano1999,Furlanetto2010}.
We use the fitting formula of \citet{Shull1985}: $x_e=n({\rm H^+})/n_{\rm H}$, where $n_{\rm H}$ is the total number density of hydrogen nuclei. 
For the predominantly neutral gas considered here, $f_{\rm ion}\simeq 0.38$. The resulting secondary yield is $Y_{\rm sec}\sim2$ at E = 0.2~keV, $\sim10$ at 1~keV, and $\sim50$ at 5~keV, showing that secondary ionisations dominate the total yield at X-ray energies.
However, the contribution of these photons to the total ionisation rate also depends on the quickly declining photoabsorption cross-section, so higher-energy photons do not necessarily dominate $\zeta_X$ despite their larger secondary yield.

The total yield is then:
\begin{equation}
Y(E)=1+Y_{\rm sec}(E).
\end{equation}

\subsection{Estimate of the deabsorbed fluxes}
The surface brightnesses derived from \textit{eROSITA} data correspond to the observed diffuse X-ray radiation field in each sky pixel ($I_{{\rm band},i }$) and are attenuated in the ISM. 
They therefore do not directly represent the radiation field incident on interstellar gas.

To estimate the intrinsic intensity, we correct for foreground absorption along the LOS.
For each sky pixel $i$, we use the corresponding total hydrogen column density $N_{{\rm H,LOS},i}$ derived from the HI4PI survey \citep{HI4PI2016}.
Table~\ref{tab:nh_los} summarises the median and 16th--84th percentile range of the pixel-level $N_{\rm H,LOS}$ distributions in each Galactic region.

For each energy band, we compute an energy-weighted transmission factor for each LOS column:
\begin{equation}
T_{\rm band}(N_{{\rm H,LOS},i}) = \frac{\int_{E_{\min}}^{E_{\max}} E^{1-\Gamma} \exp[-\sigma_{\rm abs}(E)N_{{\rm H,LOS},i}] \,\mathrm{d}E}
{\int_{E_{\min}}^{E_{\max}} E^{1-\Gamma}\,\mathrm{d}E},
\end{equation}
which corresponds to the fraction of the intrinsic band-integrated intensity transmitted through that LOS for an assumed power-law spectrum $I_E(E)\propto E^{1-\Gamma}$.

The main limitation of this procedure is the use of HI4PI-based $N_{{\rm H,LOS},i}$ as a proxy for the total foreground column.
Because additional molecular and ionised gas are not included in $N_{{\rm H,LOS},i}$, the true attenuation may be larger and the inferred intrinsic intensity correspondingly higher.

In addition to foreground absorption, part of the observed diffuse emission may originate locally (e.g., from the Local Hot Bubble or nearby hot gas) and therefore not experience the full LOS attenuation.
We account for this using a band-dependent parameter $f_{\rm back}$, defined as the fraction of the observed emission originating behind the absorbing column. The adopted values are listed in Table~\ref{tab:transm_values}.
The observed intensity in each pixel can then be written as a combination of unabsorbed foreground emission and attenuated background emission:
\begin{equation}
I_{{\rm band},i}=(1-f_{\rm back})I_{{\rm deabs},i} + f_{\rm back}I_{{\rm deabs},i} T_{\rm band}(N_{{\rm H,LOS},i}),
\end{equation}
where $I_{{\rm deabs},i}$ is the deabsorbed band-integrated intensity associated with that LOS. 
Solving for the deabsorbed intensity gives:
\begin{equation}
I_{{\rm deabs},i}= \frac{I_{{\rm band},i}}
{(1-f_{\rm back})+ f_{\rm back}T_{\rm band}(N_{{\rm H,LOS},i})}.
\end{equation}

In practice, we evaluate this correction using the pixel-level $N_{\rm H,LOS}$ distribution within each region, grouped into 400 bins of $N_{\rm H,LOS}$ for computational efficiency, and then average the corresponding deabsorbed ionisation profiles.
This procedure corrects for large-scale Galactic foreground absorption, while attenuation within MCs is treated separately in the ionisation calculations.

\section{Results}\label{sec:results}
\begin{figure}
    \centering
    \includegraphics[width=\linewidth]{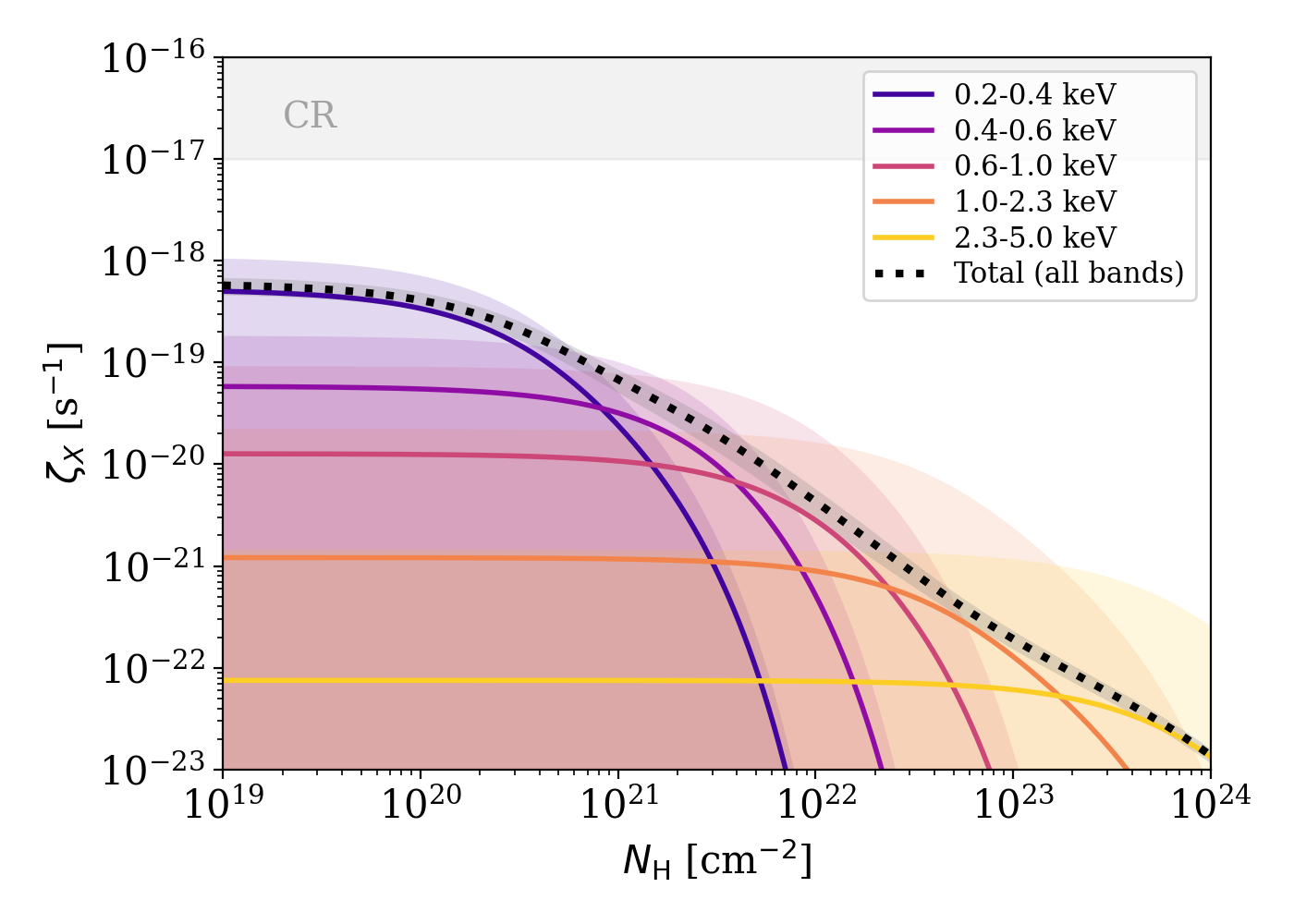}
    \caption{X-ray ionisation rate ($\zeta_{\mathrm{X}}$) as a function of hydrogen total column density ($N_{\mathrm{H}}$). Coloured curves show region-averaged contribution of each \textit{eROSITA} energy band, calculated from the observed (i.e., non-deabsorbed) diffuse surface brightness, the dotted black curve shows the total ionisation rate summed over all bands. The shaded envelope reflects the scatter across the five Galactic regions considered. The shaded grey region indicates the typical range of CRIR.}
    \label{fig:zeta_vs_NH_general}
\end{figure}

\subsection{Ionisation rate from the observed diffuse field}\label{sec:results_observed}

The results of the calculation of the X-ray ionisation rate, $\zeta_X(N_{\rm H})$ in five broad energy bands, without correcting for foreground Galactic absorption along the observer's LOS, are shown in Fig.~\ref{fig:zeta_vs_NH_general}.
Coloured curves show the mean contribution of each band across the five Galactic regions, while the dotted black curve gives the total ionisation rate summed over all bands.
As the diffuse surface brightness varies substantially within each Galactic region (see Appendix~\ref{appendix:gal_regions}, Figs.~\ref{fig:zeta_vs_NH_reg1} -- \ref{fig:zeta_vs_NH_reg5} for details), instead of using the formal statistical uncertainty on the mean, 
we quantify the intrinsic variability of the radiation field using the standard deviation of the pixel surface-brightness distribution within each region.
This scatter provides an empirical measure of the large-scale spatial variability of the diffuse X-ray emission.
The attenuation behaviour in Fig.~\ref{fig:zeta_vs_NH_general} demonstrates the steep energy dependence of the photoelectric absorption cross-section, with the softer X-ray bands being suppressed at much lower column densities than the harder bands.
The 0.2--0.4~keV band dominates $\zeta_X$ at low columns ($N_{\rm H}\lesssim10^{20}\,\mathrm{cm^{-2}}$) but declines rapidly with increasing shielding. 
At $N_{\rm H}\sim10^{21}$--$10^{22}\,\mathrm{cm^{-2}}$, the dominant contribution shifts towards harder bands ($E\gtrsim1$~keV), which remain effective up to $N_{\rm H}\sim10^{23}$--$10^{24}\,\mathrm{cm^{-2}}$. 
Beyond this depth, even the 2.3--5.0~keV band is attenuated, and the total  $\zeta_X$ continues to decline towards higher column densities.

Overall, the $\zeta_X(N_{\rm H})$ curves exhibit three regimes: (i) an optically thin plateau at low columns, 
(ii) a decline as successively softer X-ray bands are absorbed with increasing $N_{\rm H}$, and (iii) a high-column tail in which the total ionisation rate is increasingly dominated by the hardest X-ray photons and extends to $N_{\rm H}\sim10^{23}$$-$$10^{24}\,\mathrm{cm^{-2}}$.
This behaviour is consistent with classical XDR studies \citep[e.g.][]{Maloney1996,Igea1999}.

Across the explored column density range, the diffuse X-ray contribution remains below the canonical Galactic CRIR ($\zeta_{\rm CR}\sim10^{-17}-10^{-16}\,\mathrm{s^{-1}}$), showing that CRs dominate the ionisation balance.

\subsection{Deabsorbed diffuse field}\label{sec:results_deabsorbed}

\begin{table}
\caption{Hydrogen column densities in the five Galactic regions.}
\centering
\begin{tabular}{lcc}
\hline\hline
Region & $N_{\rm H,LOS}$ & Range (p16--p84) \\
 & ($10^{20}$ cm$^{-2}$) & ($10^{20}$ cm$^{-2}$) \\
\hline
I   & 30.1 & 14.5--74.2 \\
II  & 6.71 & 2.79--10.8 \\
III & 5.24 & 2.03--10.9 \\
IV  & 4.34 & 2.44--6.62 \\
V   & 3.92 & 1.87--11.4 \\
\hline
\end{tabular}
\label{tab:nh_los}
\end{table}

\begin{table}
\centering
\caption{Adopted fiducial background fractions $f_{\rm back}$ and 16th--84th percentiles ranges of pixel-level transmission $T_{\rm band}$.}
\begin{tabular}{lcc}
\hline
\hline
Energy band & $f_{\rm back}$ & $T_{\rm band}$\\
(keV) &  &  \\
\hline
0.2--0.4 & 0.1 & $2.7\times10^{-2}$ -- 0.35 \\
0.4--0.6 & 0.6 & 0.26 -- 0.79 \\
0.6--1.0 & 0.8 & 0.63 -- 0.94 \\
1.0--2.3 & 0.9 & 0.93 -- 0.99 \\
2.3--5.0 & 1.0 & 0.994 -- 0.999 \\
\hline
\end{tabular}
\label{tab:transm_values}
\end{table}

\begin{figure}
    \centering
    \includegraphics[width=\linewidth]{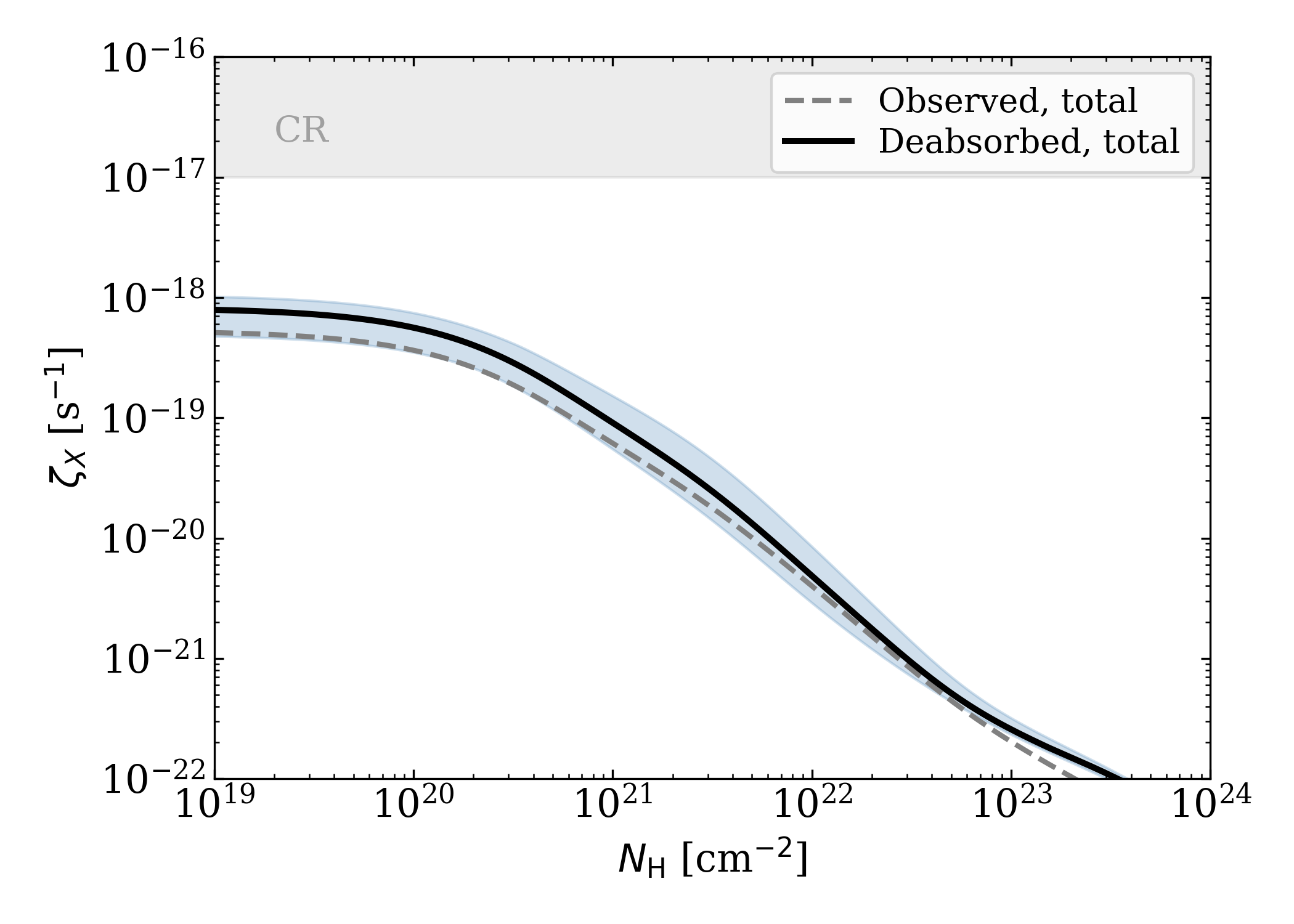}
    \caption{$\zeta_X(N_{\rm H})$ produced by deabsorbed diffuse Galactic X-ray radiation field from the N$_{\rm H}$-binned pixel-wise deabsorption procedure and summed over all \textit{eROSITA} energy bands in black and the total absorbed one in grey dashed line for comparison. 
    The blue shaded envelope represents the spread across the five Galactic regions, after propagating the pixel-level distributions of surface brightness and LOS column density.
    At $N_{\mathrm{H}} \gtrsim 10^{21} \mathrm{cm^{-2}}$ X-rays are strongly attenuated and contribute negligibly compared to CRs.}
    \label{fig:zeta_total_deabs}
\end{figure}

Next, we correct the observed surface brightness for foreground Galactic absorption along the LOS using $N_{\rm H,LOS}$ from HI4PI (see Table~\ref{tab:nh_los}) and the geometry parameter $f_{\rm back}$ (see Sect.~\ref{sec:method}). 
In practice, the deabsorption is performed on the pixel-level radiation field, grouped into bins of $N_{\rm H,LOS}$ for computational efficiency, and the resulting deabsorbed ionisation profiles are then averaged within each Galactic region.
This procedure estimates the intrinsic diffuse X-ray radiation field incident on a MC surface, while attenuation within the cloud itself is retained through the factor $\exp[-\sigma_{\rm abs}(E)N_{\rm H}]$ in the ionisation integral.
The correction mainly affects the softest X-ray bands, where even modest foreground columns strongly suppress the observed surface brightness, while harder bands are less affected.
The uncertainty envelope is obtained by propagating the regional spread in diffuse surface brightness, the pixel-level distribution of $N_{\rm H,LOS}$, and the adopted range of $f_{\rm back}$ (see Sect.~\ref{sec:discussion_NH_fbac} for discussion).
Fig.~\ref{fig:zeta_total_deabs} shows the total hydrogen ionisation rate derived from the deabsorbed diffuse radiation field, summed over all energy bands.
The black solid curve shows the median deabsorbed ionisation rate across the five Galactic regions, while the shaded envelope indicates the corresponding spread.
Correcting for foreground absorption increases the ionisation rate in optically thin gas compared to the observed (absorbed) field, but does not change the overall attenuation behaviour within the MC.
Even after deabsorption, the diffuse X-ray field produces only around $10^{-18}\,\mathrm{s^{-1}}$ ionisation rates in optically thin gas, remaining below the typical Galactic CRIR range.
\begin{figure}
    \centering
    \includegraphics[width=\linewidth]{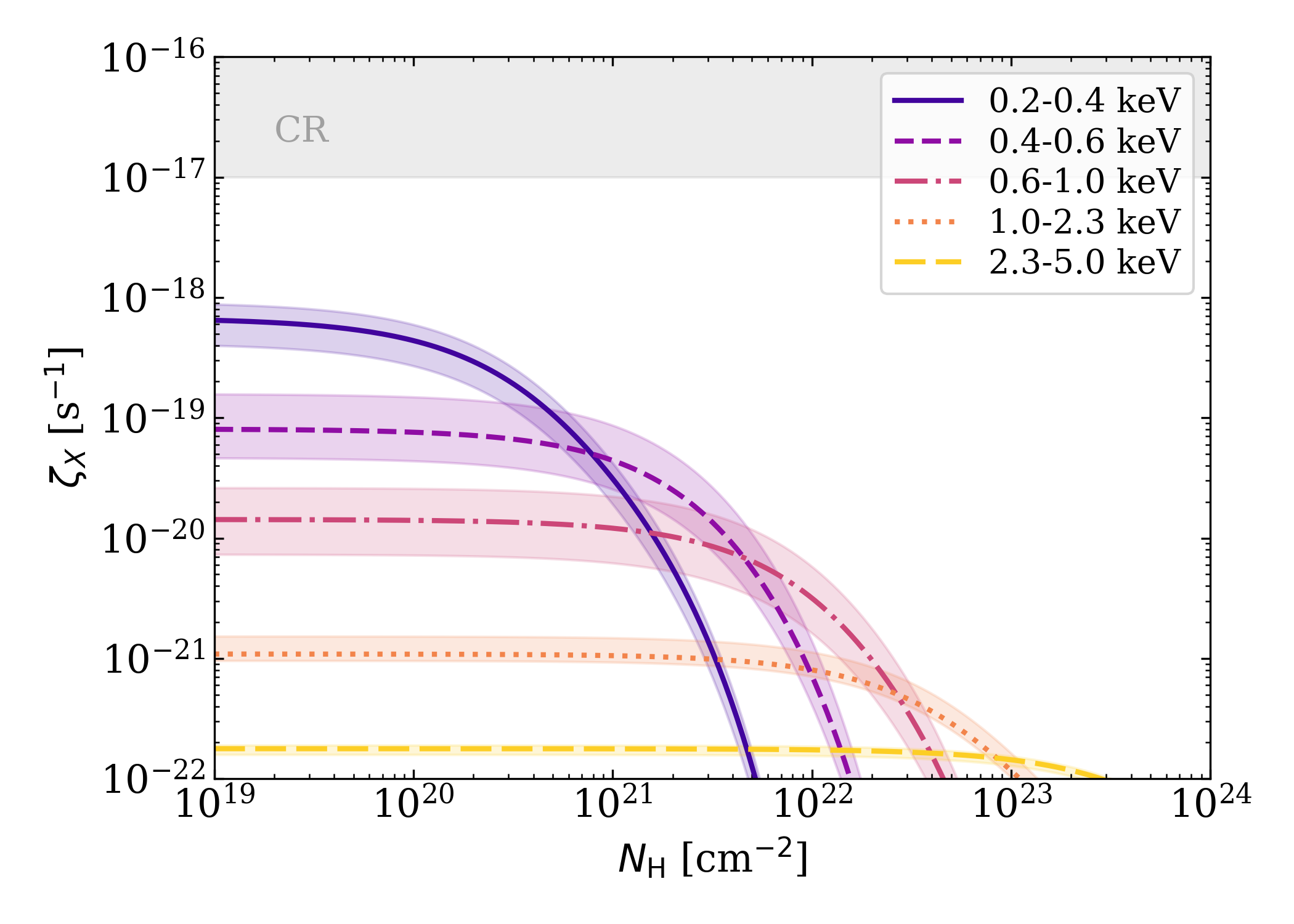}
    \caption{Contribution of individual \textit{eROSITA} energy bands to the deabsorbed diffuse X-ray ionisation rate derived from the N$_{\rm H}$-binned pixel-wise deabsorption procedure. Shaded regions are propagated uncertainty across the five Galactic regions of the pixel-level distributions of surface brightness and LOS absorption.}
    \label{fig:zeta_bands_deabs}
\end{figure}

Figure~\ref{fig:zeta_bands_deabs} shows the contributions of individual \textit{eROSITA} energy bands to the deabsorbed diffuse X-ray ionisation rate.
Each curve represents the deabsorbed radiation field within a single band, averaged over the five Galactic regions after applying the $N_{\rm H}$-binned pixel-wise deabsorption procedure, with shaded regions showing the corresponding spread.
The dominant ionising photons depend strongly on cloud column density.
At low column densities ($N_{\rm H}\lesssim10^{20}\,\mathrm{cm^{-2}}$), the ionisation rate is dominated by soft X-rays ($E\lesssim0.6$~keV),
which have the largest intensity but are rapidly absorbed. 
As shielding increases, the dominant contribution shifts progressively toward harder photons.
The 1--2.3~keV band remains up to $N_{\rm H}\sim10^{23}\,\mathrm{cm^{-2}}$, while the hardest band (2.3--5.0~keV) can influence the most shielded (deepest) cloud material.
\subsection{Spatial variation of the intrinsic field}

\begin{figure}
    \centering
    \includegraphics[width=\linewidth]{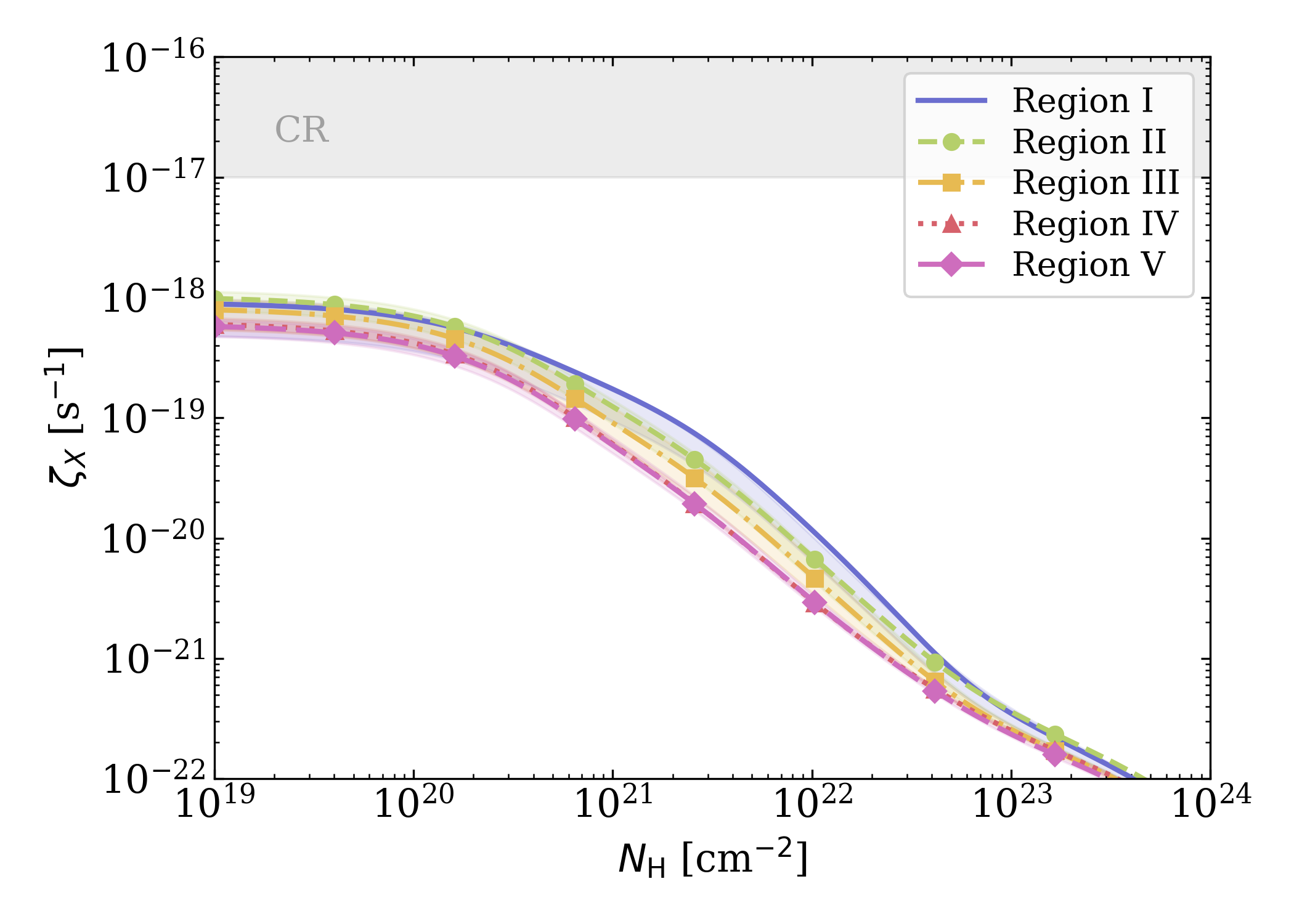}
    \caption{Spatial variation of the deabsorbed diffuse X-ray ionisation rate across the five representative Galactic regions. Each curve shows the median deabsorbed ionisation rate obtained for one region from the $N_{\rm H}$-binned pixel-wise deabsorption procedure, and shaded envelopes are the propagated spread from the pixel-level surface-brightness and $N_{\rm H,LOS}$ distributions. All regions exhibit similar attenuation behaviour with increasing column density.}
    \label{fig:zeta_regions_deabs}
\end{figure}

Figure~\ref{fig:zeta_regions_deabs} shows the variation of the deabsorbed ionisation rate across the five representative Galactic regions (see Table~\ref{tab:gal_regions}).
Each curve corresponds to the median ionisation profile obtained for one region after $N_{\rm H}$-binned pixel-wise deabsorption, with shaded envelopes representing the propagated uncertainties as above.
Despite the significant variation of surface brightness across the sky, the resulting deabsorbed ionisation rates differ only modestly between regions (typically by tens of per cent and by less than a factor of \(\sim2\) over the plotted column-density range).
All regions exhibit nearly identical attenuation behaviour with increasing column density, reflecting their broadly similar spectral energy distribution together with the universal form of the photoelectric cross-section adopted in the calculations.
The remaining differences in normalisation represent large-scale anisotropy of the diffuse Galactic X-ray background, including variations in hot plasma emission and Galactic structure.
However, these spatial variations are small compared to the attenuation produced by shielding within MCs. 
As a result, once foreground absorption is removed, the diffuse Galactic X-ray radiation field appears comparatively homogeneous on large angular scales for estimating MC ionisation rates.

\section{Discussion}\label{sec:discussion}
\subsection{Sensitivity to the assumed spectral slope}\label{sec:discussion_index}
The intrinsic diffuse spectrum within each energy band was reconstructed assuming a power-law photon index $\Gamma=2$.
Varying the spectral slope primarily redistributes energy within the band, changing the relative weighting of soft and hard photons in the attenuation integral.
Because the photoelectric cross-section decreases steeply with energy, harder spectra ($\Gamma<2$) increase the relative importance of deeply penetrating photons, while softer spectra enhance absorption at low column densities.
In practice, however, the integrated ionisation rate depends mainly on the band integrated intensity, and only then on the assumed spectral shape used to reconstruct $\mathbf{J_E(E)}$.
For possible variations of $\Gamma$ within the range typically assumed for diffuse thermal or unresolved extragalactic emission ($\Gamma \sim 1.5$--2.5) \citep{Yeung2024}, the resulting $\zeta_X(N_{\rm H})$ curves shift only modestly in normalisation, showing nearly the same attenuation behaviour.
The qualitative conclusions of this work are therefore insensitive to the exact choice of spectral index.

\subsection{Angular resolution and unresolved emission}
\begin{figure}
\centering
\includegraphics[width=\linewidth]{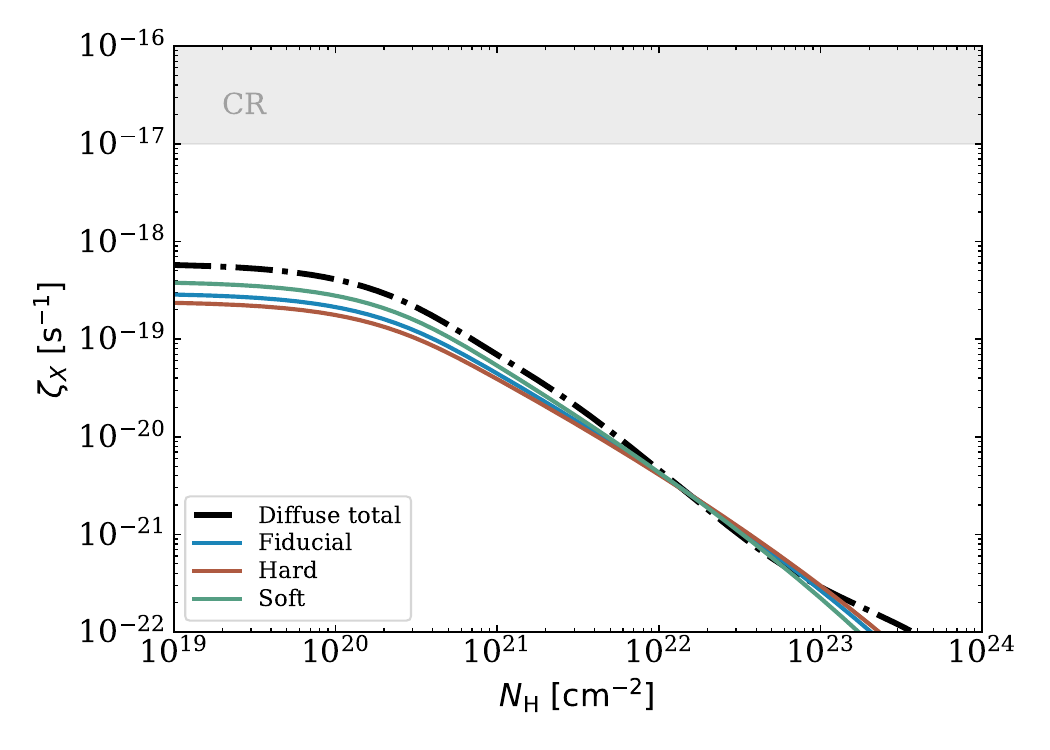}
    \caption{$\zeta_X$ as a function of column density for the diffuse \textit{eROSITA} field (black) and for unresolved-stellar radiation fields based on multi-temperature coronal emission models (coloured lines). The stellar spectra are constructed using combinations of thermal components representative of nearby FGKM stars, motivated by recent \textit{eROSITA}-derived average stellar spectra \citep{Zheng2026}, and are normalised to reproduce the same $0.2$--$2.3$ keV surface brightness as the diffuse model. Variations in the assumed stellar spectral shape (soft, fiducial, and hard cases) have only a minor impact on the resulting attenuation profiles.}
\label{fig:stellar_comparison}
\end{figure}

The diffuse maps in this work have an effective angular resolution of a few arcminutes.
At this resolution, compact sources below the point-source detection and removal threshold remain unresolved.
As a result, a non-negligible fraction of the measured diffuse emission may originate from this unresolved population of faint X-ray sources (e.g. coronally active stars, cataclysmic variables, or distant AGNs).
Recent analyses of \textit{eROSITA} data \citep{Ponti2026} suggest that the kT~$\sim0.7$ keV component of the soft X-ray background, particularly near the Galactic plane, may contain a substantial contribution from unresolved low-mass stellar populations.
To better understand the role of unresolved stellar emission, we adopt a stellar radiation field based on \textit{eROSITA}-derived average spectra of nearby FGKM stars \citep{Zheng2026}.
They are approximated as multi-temperature coronal-plasma models, combining representative M-dwarf and FGK components, and including soft, fiducial, and harder variants.
To evaluate the impact of the spectral shape, the stellar spectra are normalised to the same incident $0.2$--$2.3$ keV band integrated radiation field as the diffuse field model and propagated through MCs using the same attenuation formalism as above.
The resulting ionisation profiles $\zeta_X(N_{\mathrm H})$ are largely insensitive to the assumed stellar spectral shape as shown in Fig.~\ref{fig:stellar_comparison}.
All stellar cases show slightly lower ionisation rates and stronger attenuation at high column densities, representing the overall softness of coronal emission.
Although unresolved stellar populations may contribute significantly to the large-scale soft X-ray surface brightness, their contribution to ionisation in well-shielded MC interiors is smaller than that inferred from the diffuse-field estimate.
For this reason, the diffuse-field model provides a slightly harder and more penetrating radiation field than the unresolved-stellar cases considered here, and can therefore be taken as an upper estimate for the contribution of unresolved stellar emission to the X-ray ionisation rate.
But on smaller spatial scales (sub-pc to pc), the X-ray radiation field is more structured than captured in this work, as local sources may produce spatially inhomogeneous ionisation.

\subsection{Emission geometry and foreground components}\label{sec:discussion_NH_fbac}
The deabsorption correction depends on the assumed fraction of the observed emission originating behind the absorbing Galactic column, parameterised in this work by $f_{\rm back}$. 
This parameter reflects the uncertain spatial distribution of the diffuse X-ray emitting plasma, including contributions from local foreground components, the Galactic halo, the unresolved cosmic X-ray background, etc. 
In our fiducial model, we adopt band-dependent values of $f_{\rm back}$ (see Table~\ref{tab:transm_values} for values), motivated by the assumption that the softest emission contains mainly local contribution, while the harder bands are dominated by more distant Galactic and extragalactic components.
To test the sensitivity of our results, we also considered a higher background case with $ f_{\rm back}=(0.3,\,0.8,\,1.0,\,1.0,\,1.0)$
for the 0.2--0.4, 0.4--0.6, 0.6--1.0, 1.0--2.3, and 2.3--5.0~keV bands, correspondingly.
Relative to the fiducial model, this increases the total deabsorbed X-ray ionisation rate by only $\sim20$--30\% at $N_{\rm H}\sim10^{19-20}\,\mathrm{cm^{-2}}$, by about a factor of $\sim1.5$ around $N_{\rm H}\sim10^{21-22}\,\mathrm{cm^{-2}}$, and by at most a factor of $\sim2$ at the highest columns.
The largest differences happen at low and intermediate columns, where the softer bands contribute the most to the total ionisation rate.

We do not adopt the extreme case $f_{\rm back} = 1$  in all energy bands as a possible upper limit, assuming all of the emission has to lie behind the full Galactic absorbing column.
Because in this case, the transmission in the 0.2–-0.4~keV energy band can be as low as $T_{\rm band}\sim10^{-3}$ and the deabsorbed intensity scales as $I_{\rm deabs}=I_{\rm band}/T_{\rm band}$, resulting in correction factors of up to $\sim10^3$. 
This is an unphysical assumption that would give an enormous boost to the softest energy band.
Our high-background case provides a more realistic upper-end estimate of the X-ray ionisation rate. 
But even in this case, the X-ray ionisation rate remains below the Galactic CRIR for the well-shielded molecular gas, and our main conclusion is unchanged: diffuse X-rays are most relevant in diffuse low-column envelopes, whereas CRs dominate the ionisation of MC interiors.

In addition, our deabsorption procedure still approximates the LOS attenuation in a simplified way. 
Although the correction is applied pixel-by-pixel using the corresponding HI4PI column density $N_{{\rm H,LOS},i}$, each sightline is still treated as if the observed diffuse X-ray emission were located behind a single effective foreground column. 
In reality, the observed intensity is an integral over emitting plasma distributed continuously along the LOS, with different emitting components attenuated by various foreground columns. 
A more realistic treatment of the deabsorption would account for the distribution of such emitting plasma and optical depths along each LOS, rather than approximating the observed emission with a single foreground column per pixel. 
For example, methods of this kind have been discussed by \citet{Locatelli2022}.
Such an approach would mainly affect the deabsorbed normalisation of the softest bands, where the attenuation is most nonlinear, but is not expected to qualitatively change our conclusion that diffuse X-rays remain subdominant to CR ionisation in well-shielded MC interiors.

\subsection{Potential implications for MC chemistry}
The additional ionisation provided by diffuse X-rays at low column densities may have important consequences for the chemistry of MC envelopes and diffuse gas.
In particular, species such as Ar$^+$ and ArH$^+$ are known to trace partially shielded, predominantly atomic regions and are difficult to reproduce using CRIR alone \citep{Schilke2014,Neufeld2016,Bialy2019}.
Because Ar has an ionisation potential above the H ionisation threshold, it cannot be ionised by the FUV photons that dominate classical PDRs and instead requires EUV, X-rays or energetic particles.
In this context, our results show that the X-ray ionisation rate can become non-negligible at low  $N_{\rm H}$, providing an additional source of ionisation in regions where these species are expected to form (e.g., atomic gas; \cite{Jenkins2013}).
This implies that diffuse X-rays may contribute to the formation or survival of such tracers in low-column environments, although chemical modelling would be required to quantify that.
Another possible explanation is an enhancement of the low-energy component of the CR proton spectrum, increasing the slope of the CR proton flux at low energies \citep{Padovani2009,Ivlev2015,Padovani2018} that can reproduce the ArH$^+$ abundances observed in diffuse environments \citep{Neufeld2017}.

\subsection{Comparison with previous estimates of X-ray ionisation}
Estimates of ionisation by diffuse Galactic X-rays have previously been derived using earlier soft X-ray surveys and models of the local radiation field.
For example, soft X-rays are expected to produce hydrogen ionisation rates of order $\zeta_X \sim 10^{-18}$--$10^{-17}\,\mathrm{s^{-1}}$ in low-column-density gas, quickly decreasing with increasing shielding \citep{Wolfire2003}.
\cite{Jenkins2013} showed that, even after including diffuse X-rays with other known ionising agents, the predicted ionisation remains insufficient to explain the observed ionisation of the warm neutral medium.
Similarly, theoretical X-ray dominated region (XDR) models \citep[e.g.][]{Maloney1996,Igea1999} predict strong attenuation of soft photons and limited penetration depth compared to CRs.
Our results are broadly consistent with these earlier estimates.
Using the \textit{eROSITA} all-sky survey, we calculated ionisation rates of the same order of magnitude in optically thin gas and confirm the rapid suppression of $\zeta_X$ with increasing column density.
Even after applying our deabsorption correction, the diffuse X-ray contribution remains important mainly in low-column material and does not compete with CRIR in well-shielded MC interiors.

\section{Conclusions}\label{sec:conclusions}
We use diffuse X-ray maps from the first \textit{eROSITA} all-sky survey to quantify the ionising impact of the Galactic diffuse X-ray field on MCs.

Surface brightness measurements were converted into energy intensities, deabsorbed for foreground Galactic attenuation using HI4PI LOS column densities in a pixel-by-pixel framework, and converted into hydrogen ionisation rates $\zeta_X(N_{\rm H})$ using energy-dependent photoabsorption cross-sections and a power-law spectral model.
This allowed us to derive the depth-dependent X-ray ionisation rate across a range of shielding columns ($10^{19}$--$10^{24}\,\mathrm{cm^{-2}}$) and to estimate spatial variability, as well as systematic uncertainties.

Our main conclusions are:

\begin{enumerate}
\item Across the typical range of column densities, the diffuse X-ray ionisation rate generally remains below the typical Galactic CRIR ($\zeta_{\rm CR}\sim10^{-17}-10^{-16}\,\mathrm{s^{-1}}$; \cite{Obolentseva2024, Indriolo2026}). At $N_{\rm H}\sim10^{19}$ the deabsorbed diffuse field produces $\zeta_X\sim10^{-18}\rm s^{-1}$ with the contribution decreasing rapidly towards higher columns.
CRs dominate the ionisation balance in typical MC interiors, but diffuse X-rays may contribute in low-column-density gas and cloud envelopes.

\item The diffuse X-ray radiation field exhibits significant spatial variations across the sky.
These variations represent astrophysical inhomogeneity of the large-scale X-ray background rather than statistical measurement uncertainties.
However, the inferred ionisation rates vary only modestly between the five representative Galactic regions, showing that large-scale spatial variations have a limited effect on the predicted $\zeta_X$.

\item Correcting the observed surface brightness for foreground Galactic absorption increases the inferred intrinsic radiation field, particularly in the softest energy bands.
This correction mainly affects the absolute normalisation of $\zeta_X$ and does not change its depth dependence.
For choices of the emission-geometry parameter $f_{\rm back}$ and LOS column density, the resulting uncertainty in the total X-ray ionisation rate remains modest and does not exceed a factor of two, leaving unchanged the conclusion that diffuse X-rays are subdominant to CRs in well-shielded molecular gas.
\end{enumerate}

Although the diffuse X-ray background is a subdominant ionisation and heating source in dense molecular gas, it may influence the thermal and chemical structure of low-extinction material and the outer layers of MCs.
CRIR is known to vary significantly between individual MCs (e.g. \cite{Obolentseva2024, Indriolo2026}), and in many diffuse regions the dominant ionisation mechanism cannot be clearly determined.
If the local CRIR is sufficiently low, the diffuse X-ray background may provide a non-negligible, and in some cases competitive, contribution in such environments.
Therefore, our results provide a quantitative baseline for the large-scale Galactic X-ray radiation field.
Variations in the properties and evolutionary state of local hot-gas structures (such as the Local Bubble), further suggest that the soft X-ray background may vary between different locations in the Galaxy, introducing additional spatial variations in the corresponding X-ray ionisation rate.
Future \textit{eROSITA} data releases may allow more detailed mapping of the diffuse radiation field and its spatial variations, providing further constraints on the role of diffuse X-rays in the multiphase ISM.

\begin{acknowledgements}
We are grateful to the anonymous referee for the constructive review, which improved the clarity and quality of the manuscript.
This work is based on data from eROSITA, the soft X-ray instrument aboard SRG, a joint Russian-German science mission supported by the Russian Space Agency (Roskosmos), in the interests of the Russian Academy of Sciences represented by its Space Research Institute (IKI), and the Deutsches Zentrum für Luft- und Raumfahrt (DLR). The SRG spacecraft was built by Lavochkin Association (NPOL) and its subcontractors, and is operated by NPOL with support from the Max Planck Institute for Extraterrestrial Physics (MPE). The development and construction of the eROSITA X-ray instrument was led by MPE, with contributions from the Dr. Karl Remeis Observatory Bamberg \& ECAP (FAU Erlangen-Nuernberg), the University of Hamburg Observatory, the Leibniz Institute for Astrophysics Potsdam (AIP), and the Institute for Astronomy and Astrophysics of the University of Tübingen, with the support of DLR and the Max Planck Society. The Argelander Institute for Astronomy of the University of Bonn and the Ludwig Maximilians Universität Munich also participated in the science preparation for eROSITA. 
This research was supported by the German-Israeli Foundation for Scientific Research and Development (GIF, Grant number I-1568-303.5-2024).
S.B. acknowledges support from the ISF grant number 2071540, the GIF grant number I-1568-303.7/2024, and the Alon Fellowship prize for junior faculty.
\end{acknowledgements}

\bibliographystyle{aa}
\bibliography{formatedbib}

\begin{appendix}
\section{$\zeta_X$ for different Galactic regions}\label{appendix:gal_regions}
\begin{figure}[h]
    \centering
    \includegraphics[width=\linewidth]{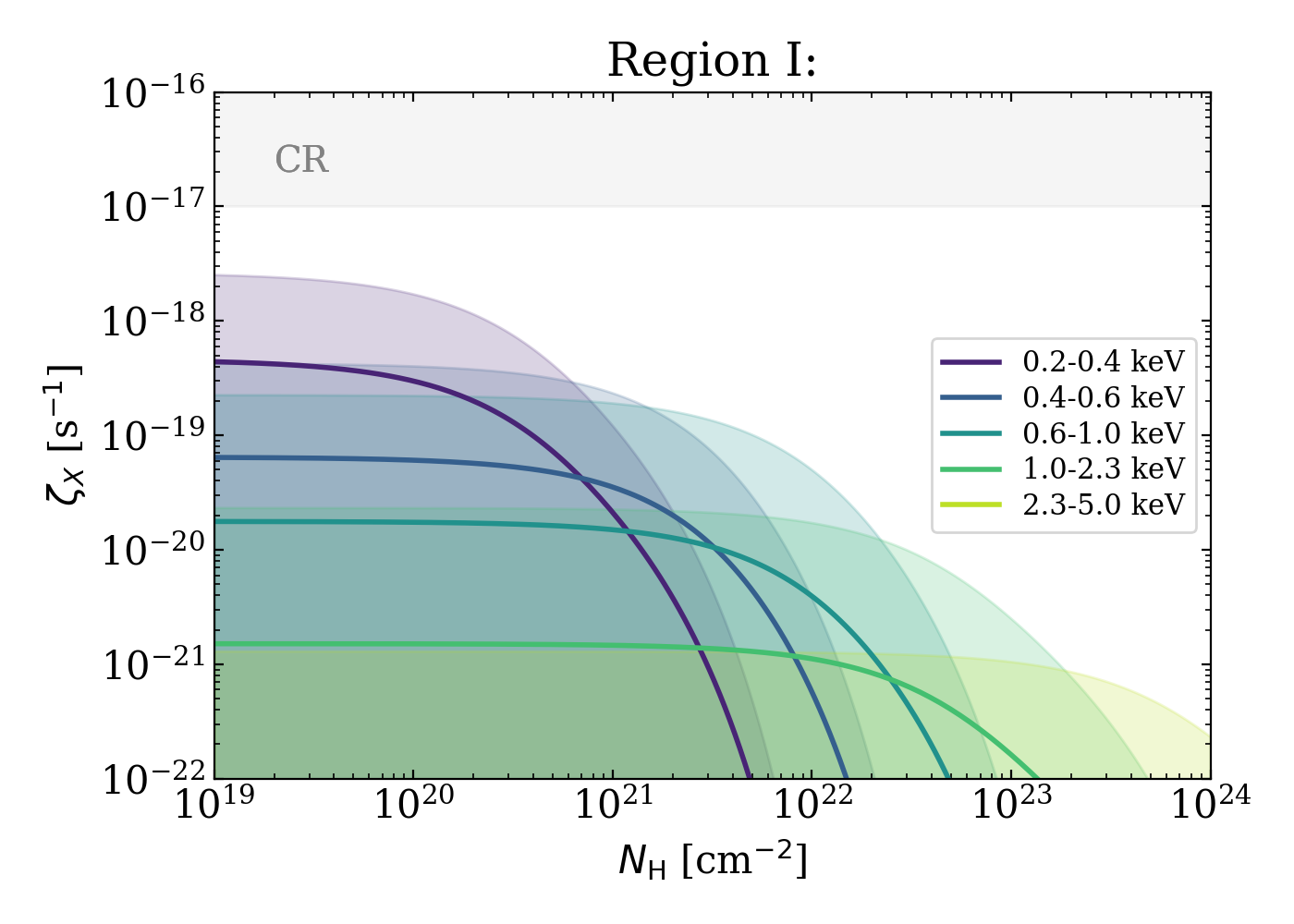}
    \caption{Same as Fig.~\ref{fig:zeta_vs_NH_general} but for Galactic region I.}
    \label{fig:zeta_vs_NH_reg1}
\end{figure}

\begin{figure}[h]
    \centering
    \includegraphics[width=\linewidth]{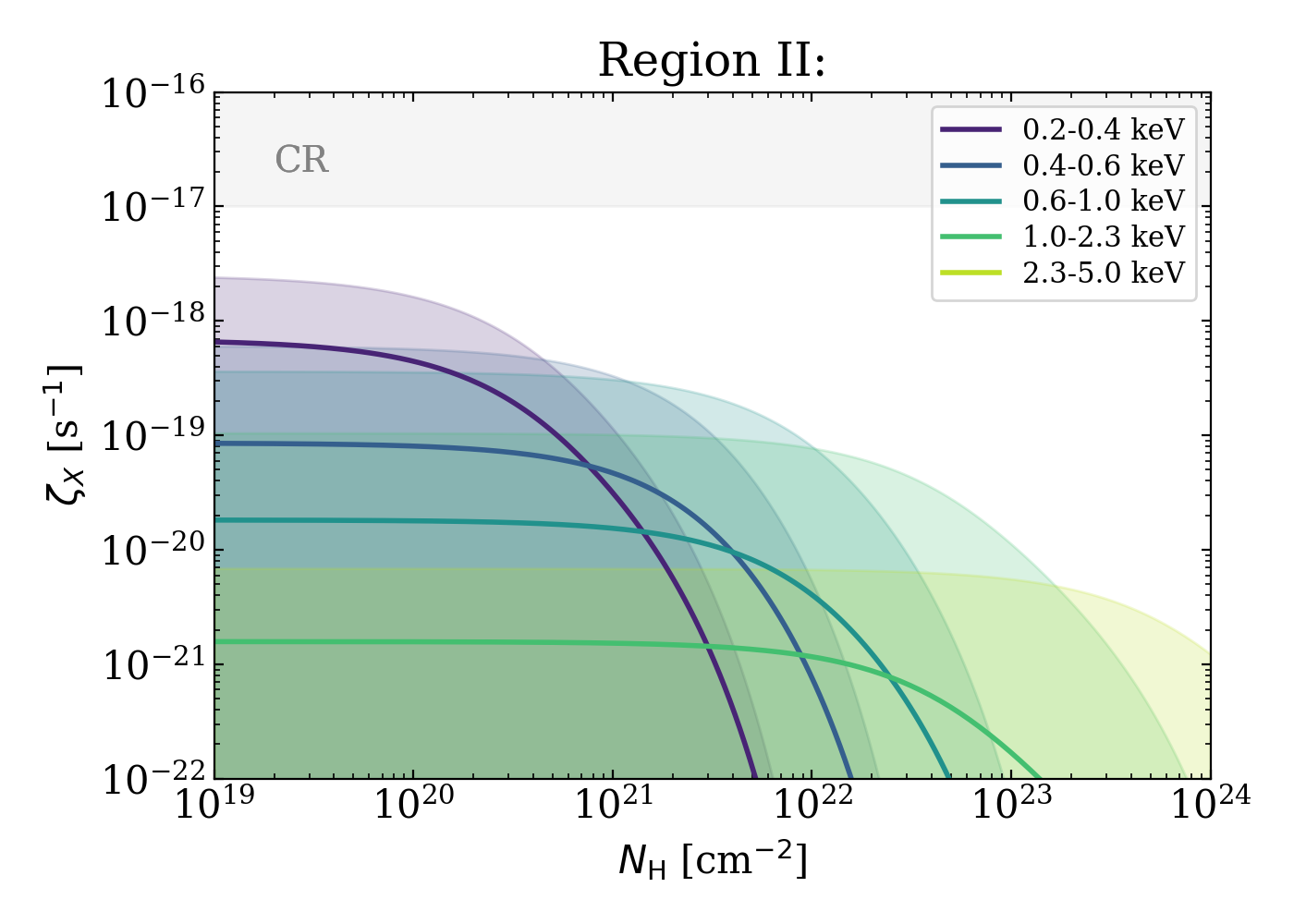}
    \caption{Same as Fig.~\ref{fig:zeta_vs_NH_general} but for Galactic region II.}
    \label{fig:zeta_vs_NH_reg2}
\end{figure}

\begin{figure}[h]
    \centering
    \includegraphics[width=\linewidth]{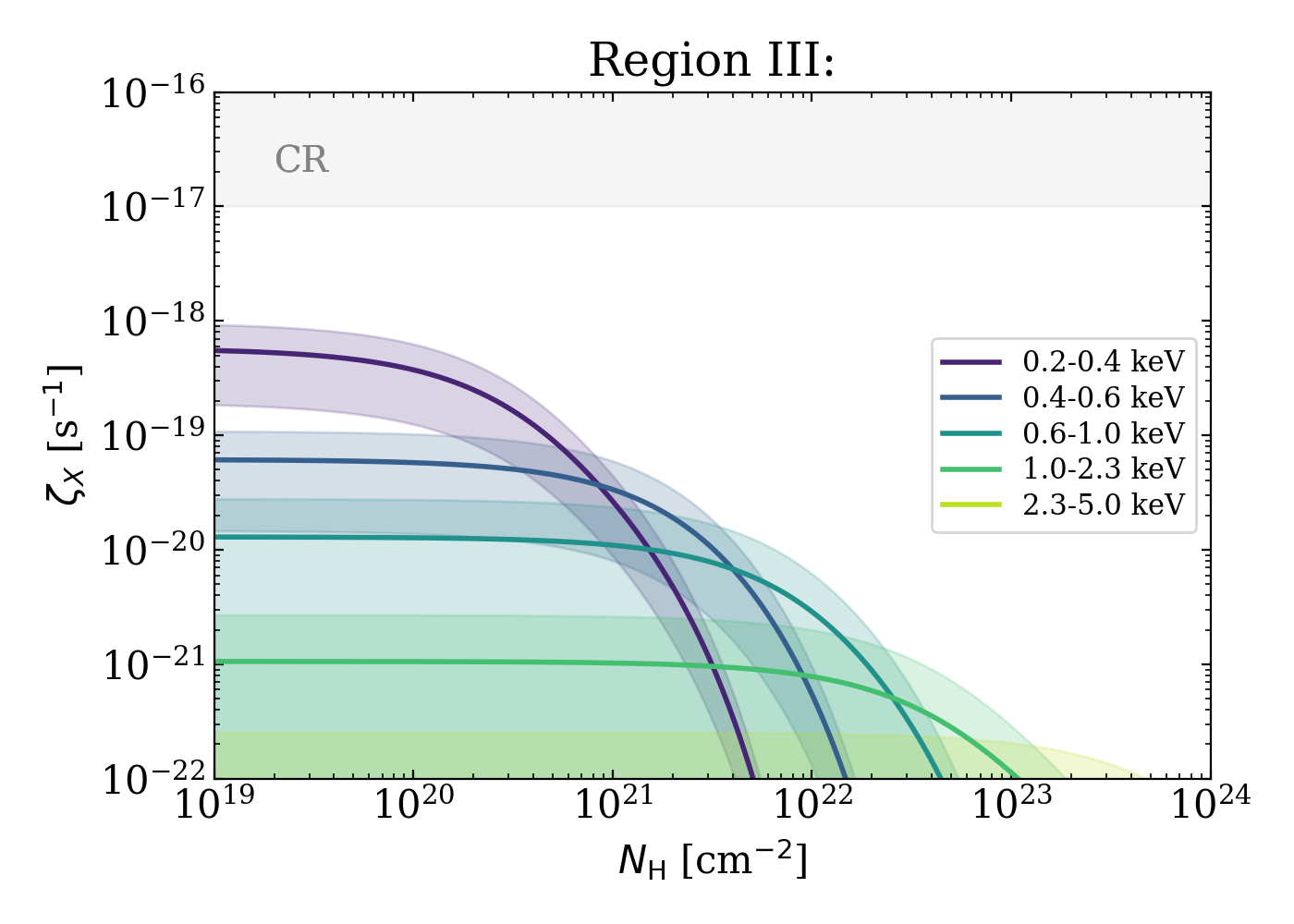}
    \caption{Same as Fig.~\ref{fig:zeta_vs_NH_general} but for Galactic region III.}
    \label{fig:zeta_vs_NH_reg3}
\end{figure}

\begin{figure}[h!]
    \centering
    \includegraphics[width=\linewidth]{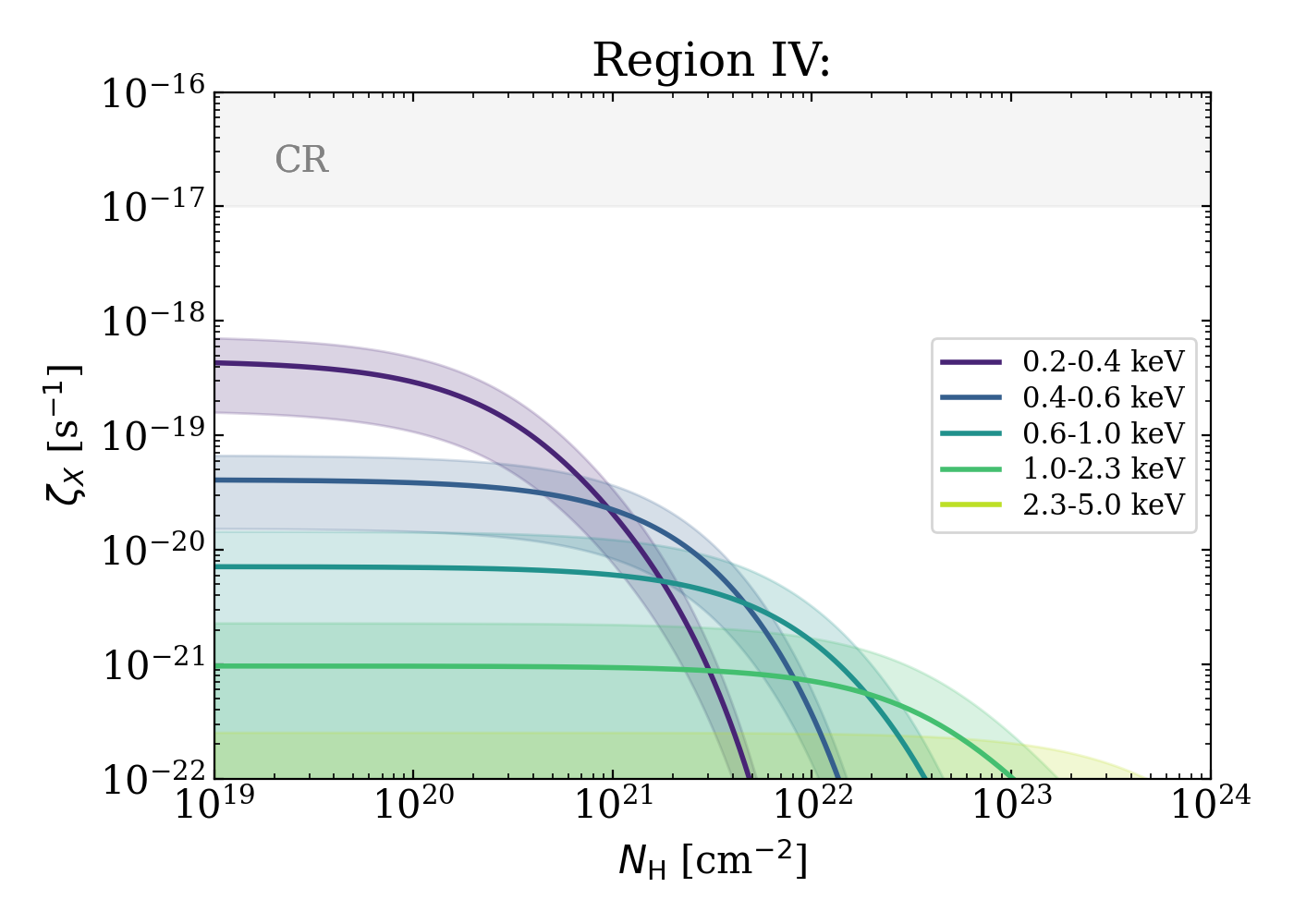}
    \caption{Same as Fig.~\ref{fig:zeta_vs_NH_general} but for Galactic region IV.}
    \label{fig:zeta_vs_NH_reg4}
\end{figure}

\begin{figure}
    \centering
    \includegraphics[width=\linewidth]{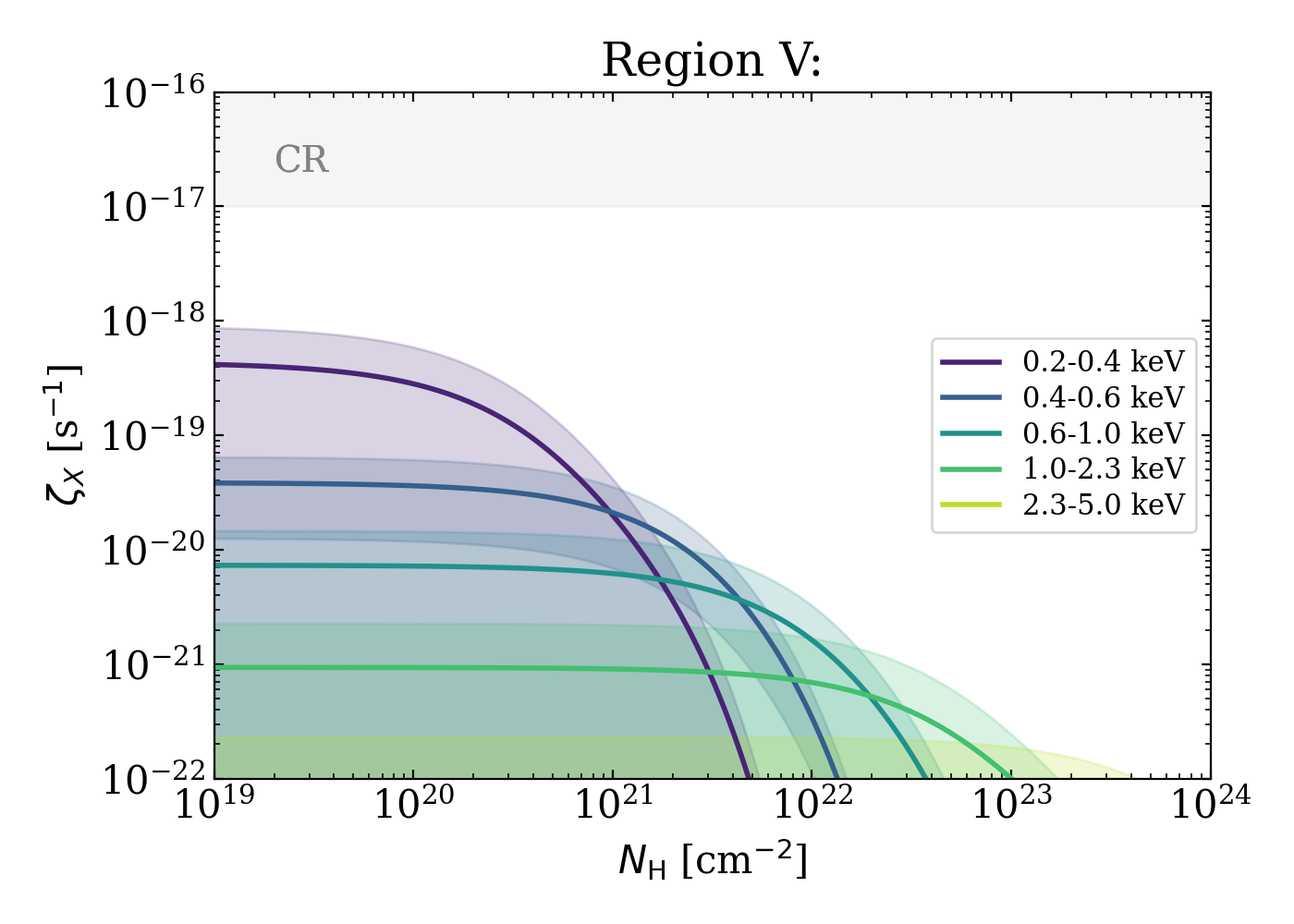}
    \caption{Same as Fig.~\ref{fig:zeta_vs_NH_general} but for Galactic region V.}
    \label{fig:zeta_vs_NH_reg5}
\end{figure}

The five Galactic regions (see Table~\ref{tab:gal_regions}) are shown in Figs.~\ref{fig:zeta_vs_NH_reg1} -- \ref{fig:zeta_vs_NH_reg5}. 
Region I samples the Galactic plane at $180^\circ < \ell < 360^\circ$ and $|b| < 15^\circ$, while Regions II–V cover higher latitudes in the outer ($180^\circ < \ell < 270^\circ$, "anti-centre") and inner ($270^\circ \leq \ell < 360^\circ$, "outflow") Galaxy, above and below the plane.
The differences between the panels show variations in the observed diffuse X-ray surface brightness across the sky. Regions at higher Galactic latitude (II–V) demonstrate systematically higher soft-band ($0.2$–$0.6$ keV) ionisation rates at low column densities compared to Region I, indicating enhanced diffuse emission away from the plane. In contrast, the Galactic-plane Region I shows a comparatively reduced soft-band contribution, but similar behaviour in the harder bands ($\gtrsim 1$ keV). At large cloud column densities ($N_{\rm H} \gtrsim 10^{22}\mathrm{cm^{-2}}$), the regional differences diminish substantially. In this regime, the ionisation rate is dominated by photons above $\sim 2$ keV, for which the observed large-scale sky variations are small compared to the strong attenuation within the MC itself.

\end{appendix}
\end{linenumbers}
\end{document}